\documentclass[aps,prd,twocolumn,showpacs,superscriptaddress]{revtex4}
\usepackage{graphics}

\begin{document}

\title{Study of the $SU(3)_c\otimes SU(3)_L\otimes U(1)_X$ model 
with the minimal scalar sector}
\author{Diego A. Guti\'errez} 
\affiliation{Instituto de F\'\i sica, Universidad de Antioquia,
A.A. 1226, Medell\'\i n, Colombia}
\author{William A. Ponce} 
\affiliation{Instituto de F\'\i sica, Universidad de Antioquia,
A.A. 1226, Medell\'\i n, Colombia}
\author{Luis A. S\'anchez}
\affiliation{Escuela de F\'\i sica, Universidad Nacional de Colombia,
A.A. 3840, Medell\'\i n, Colombia}

\begin{abstract}
A study of the three-family local gauge group
$SU(3)_c\otimes SU(3)_L\otimes U(1)_X$ with right-handed neutrinos is
carried out. We use the minimal scalar sector able to break the
symmetry in a proper way and produce, at the same time, masses for the
fermion fields. We embed the structure into a simple gauge group and, 
by using experimental results from the CERN LEP, SLAC linear collider and atomic parity violation data, we also constrain relevant parameters for the new neutral and charged currents. We discuss 
the mass spectrum for the gauge boson sector and for the spin 1/2 
particles. With the use of discrete symmetries and the introduction of 
extra scalar fields, a consistent mass spectrum could be constructed. \end{abstract}

\maketitle

\section{\label{sec:intr}Introduction}
The number of fermion families in nature and the pattern of fermion masses and mixing angles are two of the most intriguing puzzles in modern particle physics. The 3-3-1 extension of the Standard Model (SM)
of the strong and electroweak interactions, based on the
local gauge group $SU(3)_c\otimes SU(3)_L\otimes U(1)_X$, provides an interesting attempt to answer the question on family
replication. In fact, this extension has among its
best features that several models can be constructed so that anomaly
cancellation is achieved by an interplay between the families, all of them
under the condition $N_f=N_c=3$, where $N_f$ is the number of families and
$N_c$ is the number of colors of $SU(3)_c$ (three-family models)\cite{pf, vl}.

Two 3-3-1 three-family models have been studied extensively over the last
decade. In one of them the three known left-handed lepton
components for each family are associated to three $SU(3)_L$ triplets as
$(\nu_l,l^-,l^+)_L$, where $l^+_L$ is related to the right-handed isospin
singlet of the charged lepton $l^-_L$ in the SM \cite{pf}. In the other
model the three $SU(3)_L$ lepton triplets are of the form $(\nu_l, l^-,
\nu_l^c)_L$, where $\nu_l^c$ is related to the right-handed component of
the neutrino field $\nu_l$ (a model with right-handed neutrinos)\cite{vl}. 
In the first model anomaly cancellation implies quarks with
exotic electric charges $-4/3$ and $5/3$, while in the second one the extra
particles have only ordinary electric charges.

Our aim in this paper is to carry a phenomenology analysis of the 3-3-1
model in the version that includes right-handed neutrinos with a
minimal content of Higgs scalars \cite{pgs}. We are going to set updated
constraints on several parameters of the model (including those related to
the weak neutral currents), to check if the model can be embedded into a
simple gauge group structure, and most important, to check if the model
can reproduce realistic results as far as the fermion mass spectrum is
concerned.

We already know, from the analysis presented in Refs.~\cite{kita1,kita2}, that models based on the 3-3-1 local gauge structure are suitable to describe some neutrino properties, because they include in a natural way
most of the ingredients needed to explain the masses and mixing in the
neutrino sector. In particular, Ref.~\cite{kita1} addresses this issue in the context of the model studied here. Even though we are going to
concentrate mainly in other aspects of the model, we will show an alternative way to produce tiny neutrino masses with maximal mixing.

This paper is organized as follows: In Sec.~\ref{sec:sec2} we review
and present some novel features of the model. In Sec.~\ref{sec:sec3}
we study the fermion mass spectrum, and in Sec.~\ref{sec:sec4} we fix
the mixing angles between the flavor diagonal neutral currents present in
the model. In Sec.~\ref{sec:sec5} we present our conclusions.

\section{\label{sec:sec2}The model}
Some of the formulas presented in this section are taken from 
Refs.~\cite{vl} and \cite{pgs}. Corrections to some minor 
printing mistakes in the original papers are included.

\subsection{\label{sec:sec21}The gauge group} 
As it was stated above, the model we are interested in is based on the
local gauge group $SU(3)_c\otimes SU(3)_L\otimes U(1)_X$ which has 17
gauge bosons: one gauge field $B^\mu$ associated with $U(1)_X$, the 8
gluon fields $G^\mu$ associated with $SU(3)_c$ which remain massless after
breaking the symmetry, and another 8 gauge fields associated with
$SU(3)_L$ and that we write for convenience as \cite{pgs}
\begin{equation}
{1\over 2}\lambda_\alpha A^\mu_\alpha={1\over \sqrt{2}}\left(
\begin{array}{ccc}D^\mu_1 & W^{+\mu} & K^{+\mu} \\ 
W^{-\mu} & D^\mu_2 & K^{0\mu} \\
K^{-\mu} & \bar{K}^{0\mu} & D^\mu_3 \end{array}\right),
\end{equation}
where $D^\mu_1=A_3^\mu/\sqrt{2}+A_8^\mu/\sqrt{6},\;
D^\mu_2=-A_3^\mu/\sqrt{2}+A_8^\mu/\sqrt{6}$, and
$D^\mu_3=-2A_8^\mu/\sqrt{6}$. $\lambda_i, \; i=1,2,...,8$, are the eight
Gell-Mann matrices normalized as $Tr(\lambda_i\lambda_j)  
=2\delta_{ij}$.

The charge operator associated with the unbroken gauge symmetry $U(1)_Q$ 
is given by
\begin{equation}
Q=\frac{\lambda_{3L}}{2}+\frac{\lambda_{8L}}{2\sqrt{3}}+XI_3,
\end{equation}
where $I_3=Diag.(1,1,1)$ (the diagonal $3\times 3$ unit matrix), and the 
$X$ values are related to the $U(1)_X$ hypercharge and are fixed by 
anomaly cancellation. 
The sine square of the electroweak mixing angle is given by 
$S_W^2=3g_1^2/(3g_3^2+4g_1^2)$, where $g_1$ and $g_3$ are the coupling 
constants of $U(1)_X$ and $SU(3)_L$ respectively, and the photon field is 
given by
\begin{equation}\label{foton}
A_0^\mu=S_WA_3^\mu+C_W\left[\frac{T_W}{\sqrt{3}}A_8^\mu + 
\sqrt{(1-T_W^2/3)}B^\mu\right],
\end{equation}
where $C_W$ and $T_W$ are the cosine and tangent of the electroweak mixing 
angle, respectively. 

There are two weak neutral currents in the model 
associated with the two flavor diagonal neutral gauge weak bosons 
\begin{eqnarray}\nonumber \label{zzs}
Z_0^\mu&=&C_WA_3^\mu-S_W\left[\frac{T_W}{\sqrt{3}}A_8^\mu + 
\sqrt{(1-T_W^2/3)}B^\mu\right], \\ \label{zetas}
Z_0^{\prime\mu}&=&-\sqrt{(1-T_W^2/3)}A_8^\mu+\frac{T_W}{\sqrt{3}}B^\mu,
\end{eqnarray}
and one current associated with the flavor 
non diagonal neutral gauge boson $K^{0\mu}$ which carries a kind of weak 
V-isospin charge. In the former expressions 
$Z^\mu_0$ coincides with the weak neutral current of the SM. Using  
Eqs.~(\ref{foton}) and (\ref{zetas}) we can read that the gauge boson 
$Y^\mu$ associated with the $U(1)_Y$ hypercharge in the SM is 
\begin{equation} \label{hyper}
Y^\mu=\frac{T_W}{\sqrt{3}}A_8^\mu + 
\sqrt{(1-T_W^2/3)}B^\mu.
\end{equation}

\subsection{\label{sec:sec22}The spin 1/2 particle content} 
The quark content for the three families is the following:  
$Q^i_{L}=(u^i,d^i,D^i)_L\sim(3,3,0),\;i=2,3,$ for two families, where
$D^i_L$ are two extra quarks of electric charge $-1/3$ (the numbers
inside the parentheses stand for the $[SU(3)_c,SU(3)_L,U(1)_X]$ quantum
numbers in that order); $Q^1_{L}=(d^1,u^1,U)_L\sim (3,3^*,1/3)$, where
$U_L$ is an extra quark of electric charge 2/3. The right handed quarks
are $u^{ac}_{L}\sim (3^*,1,-2/3),\; d^{ac}_{L}\sim (3^*,1,1/3)$ with
$a=1,2,3,$ a family index, $D^{ic}_{L}\sim (3^*,1,1/3),\;i=2,3$, and
$U^c_L\sim (3^*,1,-2/3)$.

The lepton content is given by the three $SU(3)_L$ anti-triplets $L_{aL} =
(e_a^-,\nu_a^0,N_a^0)_L\sim (1,3^*,-1/3)$, for $a=1,2,3=e,\mu,\tau$
respectively, and the three singlets $e^+_{aL}\sim(1,1,1)$, where $\nu_a^0$
is the neutrino field associated with the lepton $e_a$ and $N_a^0$ plays
the role of the right-handed neutrino field associated to the same flavor.
Notice that this model does not contain exotic charged leptons, and
universality for the known leptons in the three families is present at
tree level in the weak basis.
 
With these quantum numbers it is just a matter of counting to check
that the model is free of the following gauge anomalies \cite{pgs}:  
$[SU(3)_c]^3$; (as in the SM $SU(3)_c$ is vectorlike); $[SU(3)_L]^3$ (six
triplets and six anti-triplets), $[SU(3)_c]^2U(1)_X; \; [SU(3)_L]^2U(1)_X
; \;[grav]^2U(1)_X$ and $[U(1)_X]^3$, where $[grav]^2U(1)_X$ stands for
the gravitational anomaly \cite{del}.

\subsection{\label{sec:sec23}$SU(6)$ as a covering group}
The Lie algebra of $SU(3)\otimes SU(3)\otimes U(1)$ is a maximal  
subalgebra of the simple algebra of $SU(6)$. The five fundamental 
irreducible representations (irreps) of $SU(6)$ are: 
$[6],[6^*],[15],[15^*]$ and the [20] which is real. The branching rules for 
these fundamental irreps into 
$SU(3)_c\otimes SU(3)_L\otimes U(1)_X$ are \cite{slansky}
\begin{eqnarray}\label{branching}\nonumber
[6] &\longrightarrow &(3,3,-1/3)\oplus (1,3,1/3), \\ \nonumber
[15] &\longrightarrow & (3^*,1,-2/3)\oplus (1,3^*,2/3) \oplus (3,3,0), \\ 
\nonumber
[20] &\longrightarrow & (1,1,1)\oplus (1,1,-1)\oplus (3,3^*,1/3) \\ \nonumber 
& & \oplus (3^*,3,-1/3),
\end{eqnarray}
where we have normalized the $U(1)_X$ hypercharge according to our needs.

From these branching rules and from the fermion structure presented above, it is clear that all the particles in this 3-3-1 model can 
be included in the following $SU(6)$ reducible representation
\begin{equation}\label{reducible}
5[6^*]+3[20]+4[15],
\end{equation}
which, besides the particles in the representations already stated 
in the previous section, includes new exotic particles, as for example
\begin{eqnarray*}
(N^0,E^+,E^{\prime +})_L&\sim&(1,3^*,2/3)\subset [15], \\
E^-_L&\sim&(1,1,-1)\subset [20], \\
(D^{\prime c},U^{\prime c},U^{\prime\prime c})L&\sim&(3^*,3,-1/3)\subset 
[20]. 
\end{eqnarray*}

The analysis shows that the reducible representation in Eq.~(\ref{reducible}) is anomalous. The simplest $SU(6)$ reducible 
representation which is free of anomalies and includes the fields in  
Eq.~(\ref{reducible}) is given by \cite{gg}
\begin{equation}\label{rred}
8[6^*]+3[20]+4[15],
\end{equation}
which includes a good deal of new exotic particles (all of them with 
ordinary electric charges): four families of 3-3-1 up and down type
quarks, four more exotic down-type quarks, plus eight families of 3-3-1 
lepton triplets, among other particles.

Our analysis, even tough producing a messy spectrum, is full of physical
content, and it seems to contradict the analysis presented in
Ref.~\cite{gg} where a three-family standard model of particles can be embedded only into a group of rank eight or larger (as for example $SU(9)$). The point here is that the extra fields in Eq.~(\ref{rred}) are not a vectorlike structure with respect to the 3-3-1 subgroup of $SU(6)$
(the structure in Eq.~(\ref{rred}) violates the so called survival
hypothesis and then the analysis presented in Refs.~\cite{gg} does not follow).

\subsection{\label{sec:sec24} The minimal scalar sector} 
For this model the minimal scalar sector able to properly break the 
symmetry, to provide with masses to the eight gauge bosons related to the 
eight broken generators in $SU(3)_L\otimes U(1)_X$ and, at the same time, 
able to provide with Yukawa terms for the fermion fields in the model, is given by \cite{pgs}: $\phi_1^T=(\phi^-_1, \phi^0_1,
\phi^{'0}_1) \sim (1,3^*,-1/3)$, and $\phi_2^T=(\phi^0_2, \phi^+_2,
\phi^{'+}_2) \sim (1,3^*,2/3)$, with Vacuum Expectation Values (VEV)
given by $\langle\phi_1\rangle^T=(0,v_1,V)$ and
$\langle\phi_2\rangle^T=(v_2,0,0)$. 

The usual analysis shows that this set of VEV breaks the symmetry in one single step
\begin{equation}
SU(3)_c\otimes SU(3)_L\otimes U(1)_X\longrightarrow SU(3)_c\otimes 
U(1)_Q.
\end{equation}
For the particular value $v_1=0$, the symmetry breaking chain becomes
\begin{eqnarray} \nonumber
3-3-1 &\stackrel{V}{\longrightarrow}& SU(3)_c\otimes SU(2)_L\otimes 
U(1)_Y \nonumber \\
& \stackrel{v_2}{\longrightarrow}& SU(3)_c\otimes U(1)_Q,
\end{eqnarray} 
which in turn allows for the matching conditions $g_2=g_3$ and 
\begin{equation}
\frac{1}{g^{\prime 2}}=\frac{1}{g_1^2}+\frac{1}{3g_2^2},
\end{equation}
where $g_2$ and $g^\prime$ are the gauge coupling constants of 
the $SU(2)_L$ and $U(1)_Y$ gauge groups in the SM, respectively. Then, for 
$v_1=0$ the SM becomes a low energy theory of this particular 3-3-1 gauge 
structure.

We will see in the next section that this scalar structure 
properly breaks the symmetry and provides with masses for the gauge 
bosons, but it is not enough to produce a consistent mass spectrum for the quark sector, and at least one more Higgs triplet, which does not develop VEV, must be introduced.

\subsection{\label{sec:sec25}The gauge boson sector}
After breaking the symmetry with $\langle\phi_i\rangle,\; i=1,2$, and
using the covariant derivative for triplets $D^\mu=\partial^\mu
-ig_3\lambda_{\alpha L}A^\mu_\alpha/2-ig_1XB_\mu I_3$, we get the 
following mass terms in the gauge boson sector.

\subsubsection{\label{sec:sec251}Spectrum in the charged gauge boson sector}
In the basis $(K^\pm_\mu, W^\pm_\mu)$, the square mass matrix produced by 
the VEV of the scalar fields is
\begin{equation} M^2_{\pm}={g^2_3 \over{2}}\left(\begin{array}{cc}
(V^2+v_2^2) & v_1V \\ v_1V & (v_1^2 + v_2^2) \end{array}\right),
\end{equation}
a symmetric mass matrix having eigenvalues
$M_{W^\prime}^2=g^2_3v_2^2/2$ and
$M_{K^\prime}^2=g^2_3(v_1^2+v_2^2+V^2)/2$, related to the physical fields
$W^\prime_\mu=\kappa (VW_\mu-v_1K_\mu),$ and $K^\prime_\mu =\kappa
(VK_\mu+v_1W_\mu)$, respectively. The first of them is associated with the known charged weak current $W^{\prime\pm}_\mu$, and the second with a new one $K^{\prime\pm}_\mu$ predicted by this model ($\kappa^{-2}=v_1^2+V^2$ is a normalization factor). From the experimental value $M_{W^\prime}=80.423 \pm 0.039$ GeV \cite{pdb} we obtain $v_2\simeq 175$ GeV as in the SM, with $v_1$ being a free parameter as far as the $M_{W^\prime}$ mass value is concerned.

Notice that for $v_1=0$ there is no mixing between the two charged gauge
bosons, and the mass for the known physical gauge boson $M_{W^\prime}^2$
is not altered. A crude value for $v_1$ can be estimated at this point in 
the following way: by using for the mass of 
$W^\prime_\mu\equiv(\cos\eta W_\mu-\sin\eta K_\mu)$ the experimental value 
given above, and using the relationship \cite{lan} 
\begin{equation} \tan^2\eta
\equiv\frac{v_1^2}{V^2}=
\frac{M_{W^\prime}^2-M_{sm}^2}{M_{K^\prime}^2-M_{W^\prime}^2} \simeq
\frac{M_{W^\prime}^2-M_{sm}^2}{g_3^2V^2/2}, 
\end{equation} 
with $M_{sm}=80.380\pm 0.023$ (the $W$ mass calculated from the SM), and $V\sim 1$ TeV, we get $0\leq\eta\leq 0.1$, which in
turn implies $0\leq v_1\leq 100$ GeV, small than the electroweak scale any
way, with the largest value for $\eta$ coming from the large experimental
and theoretical errors in the value for $M_W$. We are going to show below  
that $v_1$ must be very small since it is related to the mass scale of the neutrinos in the model.

\subsubsection{\label{sec:sec252}Spectrum in the neutral gauge boson sector}
For the five electrically neutral gauge bosons we get first, that the
imaginary part of $K^0_\mu=(K^0_{\mu R}+iK^0_{\mu I})/\sqrt{2}$ decouples
from the other four electrically neutral gauge bosons, acquiring a mass
$M^2_{K^0_I}=g^2_3(v_1^2+V^2)/2$. Then, in the basis $(B^\mu, A^\mu_3,
A^\mu_8,K^{0\mu}_R)$, a singular $4\times 4$ matrix is obtained with the 
eigenvector associated with the zero eigenvalue corresponding to the 
photon field $A_0^\mu$ in Eq.~(\ref{foton}) \cite{pgs}. The remaining $3\times 3$ mass matrix, in the basis 
$(Z^{\mu}, Z^{\prime\mu}, K^{0\mu}_R)$, reduces to
\begin{equation} \label{gauges}
\left( \begin{array}{ccc}
\frac{v_1^2+v_2^2}{\delta^2}& 
\frac{v_1^2C_{2W}-v_2^2}{\delta} & \frac{-C_W v_1 V}{\delta^2} \\
\frac{v_1^2C_{2W}-v_2^2}{\delta} & 
(v_1^2C^2_{2W}+ v_2^2+4V^2C_W^4)  &
\frac{C_W v_1V}{\delta} \\
\frac{-C_W v_1 V}{\delta^2} & \frac{C_W v_1V}{\delta}  & 
\frac{C_W^2(v_1^2+V^2)}{\delta^2} 
\end{array}\right),
\end{equation}
times a coefficient $\delta^2 g^2_3/2C_W^2$, where $C_{2W}=C^2_W-S^2_W$ 
and $\delta^{-2} = (3-4S_W^2)$. The eigenvectors and eigenvalues of this
matrix should correspond to the physical fields and their masses, 
respectively.

From this matrix we see that in the limit $v_1=0,\; K_R^{0\mu}$ does
not mix with the fields $Z^\mu$ and $Z^{\prime\mu}$ and it picks up a mass
value equal to the mass of $M_{K^0_I}$, which in turn implies that $K^{0\mu}$
and $\bar{K}^{0\mu}$ (the antiparticle of $K^{0\mu}$) have equal masses,
as they should in a well CPT behaved field theoretical framework. 

Allowing for a $v_1\neq 0$ and using the matrix entries in Eq.~(\ref{gauges}), we can read the mixing between $Z_0^\mu$ and $K^{0\mu}_R$
\begin{equation} \label{tan1} 
\tan(2\psi) = \frac{2Vv_1C_W} {V^2 C_W^2- v_2^2-v_1^2S_W^2}, 
\end{equation} 
which goes to zero in the limit $v_1=0$ as it should be. In this limit  $K^{0\mu}_R$ decouples from $Z_0^\mu$, and also from $Z_0^{\prime\mu}$. In what follows, even when $\psi\neq 0$, we will assume it takes a very small value and we will use $\cos\psi\approx 1$ and $\sin\psi\approx \psi$.

More relevant is the mixing between $Z^\mu_0$ and $Z^{\prime\mu}_0$ which 
is given by 
\begin{eqnarray} \label{tan} \nonumber
\tan(2\theta)&=& \frac{2\sqrt{(3 - 4S^2_W)}(v_2^2-v_1^2C_{2W})}
{4V^2 C_W^4 - 2v_2^2C_{2W}-v_1^2(3-4S_W^2-C_{2W}^2)}\\ 
&\stackrel{v_1=0}{\longrightarrow}&
\frac{v_2^2 \sqrt{3 - 4S^2_W}}
{2V^2 C_W^4- v_2^2C_{2W}}.
\end{eqnarray}
The physical fields are 
\begin{eqnarray}\nonumber
Z_1^\mu&=&Z^\mu_0 
\cos\theta\cos\psi-Z^{\prime\mu}\sin\theta -K^{0\mu}_R\cos\theta\sin\psi 
\; ,\\ \nonumber
Z_2^\mu&=&Z^\mu_0(\sin\theta\cos\psi\cos\phi-\sin\psi\sin\phi) 
\\ \nonumber 
& & +Z^{\prime\mu}_0 \cos\theta\cos\phi \\ \nonumber
& & -K^{0\mu}_R(\sin\phi\cos\psi+\sin\theta\sin\psi\cos\phi), 
\\ \nonumber 
K^{\prime\mu}_0&=&Z^\mu_0(\sin\theta\cos\psi\sin\phi +\sin\psi\cos\phi) 
\\ \nonumber 
& &+Z^{\prime\mu}_0 \cos\theta\sin\phi \\ \nonumber & &+K^{0\mu}_R(\cos\phi\cos\psi-\sin\theta\sin\psi\sin\phi), 
\end{eqnarray} 
where $\phi$ is the mixing angle between $Z^{\prime\mu}_0$ and 
$K^{\mu 0}_R$ which is also very small and it is zero in the limit 
$v_1=0$. In this limit the physical fields are $K^{\mu 0}$ and
\begin{eqnarray}\nonumber
Z_1^\mu&=&Z^\mu_0 
\cos\theta-Z^{\prime\mu}_0\sin\theta \; ,\\ \nonumber
Z_2^\mu&=&Z^\mu_0 \sin\theta+Z'^\mu_0 \cos\theta,  
\end{eqnarray} 
where the mixing angle $\theta$ is going to be bounded in 
Sec.~\ref{sec:sec4} using experimental constraints.

\subsection{\label{sec:sec26}Currents}
\subsubsection{\label{sec:sec261}Charged currents}
The Hamiltonian for the currents, charged under the generators of the 
$SU(3)_L$ group, is  
$H^{CC}=g_3(W^+_\mu J^\mu_{W^+}+K^+_\mu J_{K^+}^\mu+K^0_\mu 
J_{K^0}^\mu)/\sqrt{2}+h.c.$, with

\begin{eqnarray}\nonumber
J_{W^+}^\mu&=& (\sum_{i=2}^3\bar{u}^i_{L}\gamma^\mu d^i_{L})
-\bar{u}^1_{L}\gamma^\mu d^1_{L} - \sum_{a=e,\mu,\tau}
\bar{\nu}_{a L}\gamma^\mu e^-_{aL},\\ \nonumber 
J^\mu_{K^+}&=&(\sum_{i=2}^3\bar{u}^i_{L}\gamma^\mu 
D^i_{L})-\bar{U}_{L}\gamma^\mu 
d^1_{L} - \sum_{a=e,\mu,\tau}
\bar{N}^0_{a L}\gamma^\mu e^-_{aL}, \\ \nonumber 
J^\mu_{K^0}&=&(\sum_{i=2}^3\bar{d}^i_{L}\gamma^\mu 
D^i_{L})-\bar{U}_{L}\gamma^\mu u^1_{L} - \sum_{a=e,\mu,\tau}
\bar{N}^0_{a L}\gamma^\mu \nu_{a L}, \nonumber
\end{eqnarray} 
where $K^0_\mu$ is electrically neutral but carries weak V-isospin, besides 
it is flavor non diagonal.

\subsubsection{\label{sec:sec262}Neutral currents}

The neutral currents $J_\mu(EM), \; J_\mu(Z)$ and $J_\mu(Z^\prime)$,
associated with the Hamiltonian $H^0 = eA^\mu J_\mu(EM)+(g_3/{C_W})Z^\mu
J_\mu(Z)+ (g_1/\sqrt{3})Z^{\prime\mu} J_\mu(Z^\prime)$, are

\begin{eqnarray}\nonumber
J_\mu(EM)&=&{2\over 3}\left[\sum_{a=1}^3\bar{u}_a\gamma_\mu u_a +
\bar{U}\gamma_\mu U \right] \\ \nonumber 
& &- {1\over 3}\left[\sum_{a=1}^3\bar{d}^a\gamma_\mu d^a+ 
\sum_{i=1}^2\bar{D}^i\gamma_\mu D^i \right]  \\ \nonumber
& &- \sum_{a=e,\mu,\tau}\bar{e}^-_a\gamma_\mu e^-_a \\ \nonumber
& =& \sum_f q_f\bar{f}\gamma_\mu f,\\* \nonumber
J_\mu(Z)&=&J_{\mu,L}(Z)-S^2_WJ_\mu(EM),\\ \nonumber
J_\mu(Z^\prime)&=& - J_{\mu,L}(Z^\prime)+T_WJ_\mu(EM), 
\end{eqnarray}
where $e=g_3S_W=g_1C_W\sqrt{(1-T_W^2/3)}>0$ is the electric charge, 
$q_f$ is the electric charge of the fermion $f$ in units of $e$, and 
$J_\mu(EM)$ is the electromagnetic current. 

The left-handed currents are
\begin{eqnarray} \nonumber
J_{\mu,L}(Z)&=&{1\over 2}[\sum_{a=1}^3(\bar{u}^a_{L}\gamma_\mu u^a_{L}
-\bar{d}^a_{L}\gamma_\mu d^a_{L}) \\ \nonumber
& &+ \sum_{a=e,\mu,\tau}(\bar{\nu}_{a L}\gamma_\mu \nu_{a L} 
-\bar{e}^-_{a L}\gamma_\mu e^-_{a L})] \\ 
&=&\sum_F \bar{F}_LT_{3f}\gamma_\mu F_L ,
\end{eqnarray}
\begin{eqnarray}\nonumber
J_{\mu,L}(Z^\prime)&=& 
S^{-1}_{2W}[\bar{u}_{2L}\gamma_\mu u_{2L}+\bar{u}_{3L}\gamma_\mu u_{3L}
-\bar{d}_{1L}\gamma_\mu d_{1L} \\ \nonumber
& &-\sum_a(\bar{e}_{aL}\gamma_\mu e_{aL})] \\ \nonumber
& &+T^{-1}_{2W}[\bar{d}_{2L}\gamma_\mu u_{dL}+\bar{d}_{3L}\gamma_\mu 
u_{dL} \\ \nonumber
& & -\bar{u}_{1L}\gamma_\mu u_{1L}-\sum_a(\bar{\nu}_{aL}\gamma_\mu \nu_{aL})] \\  \nonumber
& &+T^{-1}_{W}[\bar{D}_{2L}\gamma_\mu D_{2L}+\bar{D}_{3L}\gamma_\mu 
D_{3L} \\ \nonumber 
& & -\bar{U}_{1L}\gamma_\mu U_{1L} 
-\sum_a(\bar{N}_{aL}\gamma_\mu N_{aL})] \\
&=&\sum_F\bar{F}_L T^\prime_{3f}\gamma_\mu F_L,
\end{eqnarray}
where $S_{2W}=2S_WC_W,\; T_{2W}=S_{2W}/C_{2W}, \;T_{3f}=Dg(1/2,-1/2,0)$ is
the third component of the weak isospin, $T^\prime_{3f}=Dg(S^{-1}_{2W},
T^{-1}_{2W}, -T_W^{-1})$ is a convenient $3\times 3$ diagonal matrix,
acting both of them on the representation 3 of $SU(3)_L$ (the negative
value when acting on the representation $3^*$, which is also true for 
the matrix $T_{3f}$) and $F$ is a generic
symbol for the representations 3 and $3^*$ of $SU(3)_L$.  Notice that
$J_\mu(Z)$ is the neutral current of the SM (with the extra fields 
included in $J_\mu (EM$)). This allows us to identify $Z_\mu$ as the 
neutral gauge boson of the SM, which is consistent with Eqs.~(\ref{zzs}) 
and (\ref{hyper}).

The couplings of the flavor diagonal mass eigenstates $Z_1^\mu$ and 
$Z_2^\mu$ are given by
\begin{eqnarray} \nonumber
H^{NC}&=&\frac{g_3}{2C_W}\sum_{i=1}^2Z_i^\mu\sum_f\{\bar{f}\gamma_\mu
[a_{iL}(f)(1-\gamma_5) \\ \nonumber
& & +a_{iR}(f)(1+\gamma_5)]f\} \\ \nonumber
      &=&\frac{g_3}{2C_W}\sum_{i=1}^2Z_i^\mu\sum_f\{\bar{f}\gamma_\mu
      [g(f)_{iV}-g(f)_{iA}\gamma_5]f\},
\end{eqnarray}
with
\begin{eqnarray} \nonumber
a_{1L}(f)&=&\cos\theta\cos\psi(T_{3f}-q_fS^2_W) \\ \nonumber
& & +\Theta\sin\theta 
(T^\prime_{3f}-q_fT_W), \\ \nonumber
a_{1R}(f)&=&-q_f\left(\cos\theta\cos\psi S_W^2
+\Theta\sin\theta T_W\right),\\ \nonumber
a_{2L}(f)&=&S_A(T_{3f}-q_fS^2_W) \\ \nonumber
& & -\Theta\cos\theta\cos\phi 
(T^\prime_{3f}-q_fT_W),\\ 
\label{a}
a_{2R}(f)&=&-q_f\left(S_A S^2_W-\Theta\cos\theta\cos\phi T_W\right),
\end{eqnarray}
where $S_A=(\sin\theta\cos\phi\cos\psi - \sin\phi\sin\psi)$, and 
$\Theta = S_WC_W/\sqrt{(3-4S_W^2)}$.
From these coefficients we can read 
\begin{eqnarray} \nonumber
g(f)_{1V}&=&\cos\theta\cos\psi(T_{3f}-2q_fS^2_W) \\ \nonumber
& & +\Theta\sin\theta (T^\prime_{3f}-2q_fT_W), \\ \nonumber
g(f)_{2V}&=&S_A(T_{3f}-2q_fS^2_W) \\ \nonumber
& & -\Theta\cos\theta\cos\phi 
(T^\prime_{3f}-2q_fT_W),\\ \nonumber
g(f)_{1A}&=&\cos\theta\cos\psi 
T_{3f}+\Theta\sin\theta T^\prime_{3f}, \\ \label{g}
g(f)_{2A}&=&S_A T_{3f}-\Theta\cos\theta\cos\phi T^\prime_{3f}. 
\end{eqnarray}
The values of $g_{iV}$ and $g_{iA}$, with $i=1,2$, are listed in Tables~\ref{tab1} and \ref{tab2}.

\begin{table*}
\caption{\label{tab1}The $Z_1^\mu\longrightarrow \bar{f}f$ couplings.}
\begin{ruledtabular}
\begin{tabular}{lcc}
$f$ & $g(f)_{1V}$ & $g(f)_{1A}$ \\ \hline
$u^{2,3}$& 
$({1\over 2}-{4S_W^2 \over 3})\cos\theta\cos\psi
+\Theta(s_{2W}^{-1}-{4T_W \over 3})\sin\theta$
& ${1\over 2}\cos\theta\cos\psi + \Theta S_{2W}^{-1}\sin\theta$  \\ 
$u^{1}$&$({1\over 2}-{4S_W^2 \over 3})\cos\theta\cos\psi -  
\Theta (T_{2W}^{-1}+{4T_W\over 3})\sin\theta$ & 
${1\over 2}\cos\theta\cos\psi - \Theta T_{2W}^{-1}\sin\theta$\\
$d^{2,3}$ & $(-{1\over 2}+{2S_W^2\over 3})\cos\theta\cos\psi 
+\Theta(T_{2W}^{-1}+{2T_W\over 3})\sin\theta$ 
& $-{1\over 2}\cos\theta\cos\psi +\Theta T_{2W}^{-1}\sin\theta$ \\
$d^{1}$ & $(-{1\over 2}+{2S_W^2\over 3})\cos\theta\cos\psi
-\Theta(S_{2W}^{-1}-{2T_W\over 3})\sin\theta$ & 
$-{1\over 2}\cos\theta\cos\psi - \Theta S_{2W}^{-1}\sin\theta$\\
$U$ & $-{4S_W^2\over 3}\cos\theta\cos\psi-\Theta(T_W^{-1}+
{4T_W\over 3})\sin\theta $ & $\Theta T_W^{-1}\sin\theta $ \\
$D_{1,2}$ & ${2S_W^2\over 3}\cos\theta\cos\psi
+\Theta (T_W^{-1}+{2T_W\over 3})\sin\theta $ &
$-\Theta T_W^{-1}\sin\theta $ \\
$e^-_{1,2,3}$& $(-{1\over 2}+2S_W^2)\cos\theta\cos\psi -
\Theta(S_{2W}^{-1}-2T_W)\sin\theta $ 
& $ -{1\over 2}\cos\theta\cos\psi-\Theta S_{2W}^{-1}\sin\theta $\\
$\nu_{1,2,3}$ & ${1\over 2}\cos\theta\cos\psi -\Theta 
T_{2W}^{-1}\sin\theta$ 
& ${1\over 2}\cos\theta\cos\psi-\Theta T_{2W}^{-1}\sin\theta$ \\
$N^0_{1,2,3}$ & $-\Theta T_W^{-1}\sin\theta $ 
& $-\Theta T_W^{-1}\sin\theta $ \\ 
\end{tabular}
\end{ruledtabular}
\end{table*}

\begin{table*}
\caption{\label{tab2}The $Z_2^\mu\longrightarrow \bar{f}f$ couplings.}
\begin{ruledtabular}
\begin{tabular}{lcc}
$f$ & $g(f)_{2V}$ & $g(f)_{2A}$ \\ \hline
$u^{2,3}$& $({1\over 2}-{4S_W^2 \over 3})S_A-\Theta (S_{2W}^{-1}-
{4T_W\over 3})\cos\theta\cos\phi $
& ${1\over 2}S_A-\Theta S_{2W}^{-1}\cos\theta\cos\phi $ \\ 
$u^{1}$&$({1\over 2}-{4S_W^2 \over 3})S_A +\Theta (T_{2W}^{-1}+ {4T_W\over 
3})\cos\theta\cos\phi $
& ${1\over 2}S_A+\Theta T_{2W}^{-1}\cos\theta\cos\phi $\\
$d^{2,3}$ & $(-{1\over 2}+{2S_W^2\over 3})S_A-
\Theta(T_{2W}^{-1}+{2T_W\over 3})\cos\theta\cos\phi $
& $-{1\over
2}S_A-\Theta T_{2W}^{-1}\cos\theta\cos\phi $ \\
$d^{1}$ & $(-{1\over 2}+{2S_W^2\over 3})S_A+
\Theta (S_{2W}^{-1}-{2T_W\over 3})\cos\theta\cos\phi$ & 
$-{1\over 2}S_A+\Theta S_{2W}^{-1}\cos\theta\cos\phi $\\
$U$ & $-{4S_W^2\over 3}S_A + 
\Theta(T_W^{-1}+{4T_W\over 3})\cos\theta\cos\phi $ & 
$\Theta T_W^{-1}\cos\theta\cos\phi $\\
$D^{2,3}$ & ${2S_W^2\over 3}S_A
-\Theta(T_W^{-1}+{2T_W\over 3})\cos\theta\cos\phi$ &
$-\Theta T_W^{-1}\cos\theta\cos\phi $ \\
$e^-_{1,2,3}$& $(-{1\over 2}+2S_W^2)S_A+\Theta(S_{2W}^{-1}-
{2T_W\over 3})\cos\theta\cos\phi $ & 
$-{1\over 2}S_A+\Theta S_{2W}^{-1}\cos\theta\cos\phi $\\
$\nu_{1,2,3}$ & ${1\over 2}S_A+\Theta T_{2W}^{-1}\cos\theta\cos\phi $ & 
${1\over 2}S_A+\Theta T_{2W}^{-1}\cos\theta\cos\phi $ \\
$N^0_{1,2,3}$ & $\Theta T_W^{-1}\cos\theta\cos\phi $ &
$\Theta T_W^{-1}\cos\theta\cos\phi $ \\
\end{tabular}
\end{ruledtabular}
\end{table*}

As we can see, in the limit $\theta =0$ the couplings of
$Z_1^\mu$ to the ordinary leptons and quarks are the same as in the SM;
due to this property we can test the new physics beyond the SM predicted 
by this particular model.

\section{\label{sec:sec3}Fermion Masses}
The Higgs scalars introduced in Sec.~\ref{sec:sec24} break the symmetry in 
an appropriate way and, at the same time, are the only scalar 
representations able to produce mass terms for the fermion fields via 
Yukawa interactions. 

\subsection{\label{sec:sec31}The up quark sector}
The Yukawa terms for the up quark sector are
\begin{eqnarray}\label{mup} \nonumber
{\cal L}^u_Y&=& Q_L^1\phi_1^*C(h^UU_L^c+\sum_{a=1}^3h_a^uu_L^{ac})\\  
&+& \sum_{i=2}^3Q^i_L\phi_2C(\sum_{a=1}^3h^u_{ia}u_L^{ac}+h_i^UU_L^c) 
+ h.c.,
\end{eqnarray}
where the $h's$ are Yukawa couplings and $C$ is the charge conjugation
operator. By introducing a symmetry which forbids the term proportional to
$h_a^u$ and $h_i^U$ (see below) and in the basis $(u^1,u^2,u^3,U)$ we get, 
from Eq.~(\ref{mup}), the following tree-level mass matrix  
\begin{equation}
M_u=\left(\begin{array}{cccc} 0 & h^{u}_{21}v_2 & h^{u}_{31}v_2 & 0 \\ 0 &
h^{u}_{22}v_2 & h^{u}_{32}v_2 & 0 \\ 0 & h^{u}_{23}v_2 & h^{u}_{33}v_2 & 0
\\ h^Uv_1 & 0 & 0 & h^UV \\ \end{array}\right), 
\end{equation} 
The matrix $M_u^\dagger M_u$ has one eigenvalue equal to zero associated to the quark $u^1$ which we identify as the $u$ quark in the first family. The other three eigenvalues are: one at the scale $V^2$ with eigenvector $U$ and two at the electroweak scale $v_2^2$, one of them suppressed by differences of Yukawa couplings; let us see:

First, if we set the 8 Yukawa couplings equal to a common real value $h^u$, then two zero eigenvalues are obtained, one corresponding to $u^1$ and the other 
one associated to $(u^2-u^3)/\sqrt{2}$. The two eigenvalues different from zero
are: $(h^uV)^2$ related to the exotic $U$ quark and $(2h^uv_2)^2$ related
to the eigenvector $(u^2+u^3)/\sqrt{2}$ which we may identify as the top
quark $t$, with a proper mass if we set $h^u=0.5$.

Next, let us set $h^u_{21}=h^u_{31}=h^U=h^u,\; h^u_{22}=h^u_{33}=h_a$ and
$h^u_{23}=h^u_{32}=h_b$, then we have now only one zero eigenvalue related
with the $u^1$ quark in the first family, one eigenvalue equal to
$(h_uV)^2$ associated to the exotic $U$ quark, a third eigenvalue
$(h_a^2+h_b^2)v_2^2$ with eigenvector $(u^2+u^3)/\sqrt{2}$
that we may identify with the top quark $t$, and a fourth eigenvalue
$(h_a^2-h_b^2)v_2^2$ with eigenvector $(u^2-u^3)/\sqrt{2}$ 
that we may identify with the charm quark $c$.

This result is quite interesting by itself because it allows us to
identify, for the up quark sector, the
mass eigenstates as a function of the weak eigenstates, and the model
becomes a realistic one (at least for the up quark sector) as far as we
can identify a mechanism able to generate afterwards a mass for the $u$ 
quark in the first family.

\begin{figure}
\includegraphics{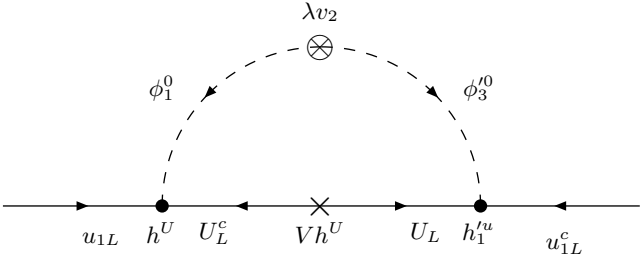}
\caption{\label{fig1}One loop diagram contributing to the radiative 
 generation of the up quark mass.}
\end{figure}

Clearly there is not an ingredient able to generate this mass in the model presented so far. What we propose here
is to introduce a third scalar $\phi_3^T=(\phi_3^-,\phi_3^0,\phi_3^{\prime
0})\sim (1,3^*,-1/3)$ which does not develop a VEV, but that couples to the
up quark sector via a Yukawa term of the form
$Q_L^1\phi_3^*C\sum_{a}h_a^{\prime u}u_L^{ac}$. Then the up quark picks
up a radiative mass via the diagram depicted in Fig.~\ref{fig1}, where the
mixing in the neutral sector is due to a totally antisymmetric term in the
scalar potential of the form $\lambda\phi_1\phi_2\phi_3$, where the
$SU(3)_L$ indexes are understood.

The restrictions on the Yukawa couplings in the up quark sector can be realized by requiring invariance under an anomaly free discrete $Z_2$ symmetry \cite{ross}, with the following assignments of $Z_2$ charge $q$
\begin{eqnarray*}\label{z2d} 
q(Q^a_{L}, U^{c}_L, D_L^{ic}, e^{ac}_L, \phi_1)&=& 0, \\ 
q(u^{ac}_{L}, d_L^{ac}, L^{a}_L, \phi_2, \phi_3)&=& 1,
\end{eqnarray*}
for $a=1,2,3$ and $i=1,2$; where we have included the down quark and 
lepton sectors, anticipating the analysis which follows.

As can be seen, we have avoided a hierarchy of the Yukawa couplings in the up quark sector just by introducing the extra Higgs field $\phi_3$, obtaining in this way a neat mass spectrum in this sector.

\subsection{\label{sec:sec32}The down quark sector}
The most general Yukawa terms for the down quark sector, using just 
$\phi_1$ and $\phi_2$, are
\begin{eqnarray}\label{mdown} \nonumber
{\cal L}^d_Y&=& 
Q_L^1\phi_2^*C(\sum_ih^D_iD_L^{ic}+\sum_{a}h_a^dd_L^{ac})\\ \nonumber
&+&\sum_{i}Q^i_L\phi_1C(\sum_ah^d_{ia}d_L^{ac}+\sum_jh^D_{ij}D_L^{jc}) \\
&+&h.c.
\end{eqnarray}
In the basis $(d^1,d^2,d^3,D^2,D^3)$ and with the $Z_2$ symmetry 
introduced above, this lagrangian produces a tree-level down quark mass 
matrix of the form
\begin{equation}
M_d=\left(\begin{array}{ccccc}
h_1^dv_2 & 0  & 0  & 0  & 0 \\
h_2^dv_2 & 0  & 0  & 0  & 0 \\
h_3^dv_2 & 0  & 0  & 0  & 0 \\
0 & h_{22}^Dv_1 & h_{32}^Dv_1 & h_{22}^DV & h_{32}^DV\\
0 & h_{23}^Dv_1 & h_{33}^Dv_1 & h_{23}^DV & h_{33}^DV\\
\end{array}\right).
\end{equation}
The matrix $M_d^\dagger M_d$ has two eigenvalues equal to zero (even for 
$v_1\neq 0$). For $h_1^d=h_2^d=h_3^d \equiv h^d$, and $h_{ij}^D$ of order 
one, the nonzero eigenvalues are: two of the order of $V$ that we may 
identify with the masses of the heavy exotic quarks $D^1$ and $D^2$, and one of the order of $3h^dv_2$ that we may identify with the mass of the ordinary down quark $d^1$, with a hierarchy $h^d/h^u\sim 10^{-4}$ (this hierarchy can be eliminated by introducing more scalars). The two states with zero eigenvalues are linear combinations of $d^2$ and $d^3$ that we may identify with the bottom and strange quarks.

\begin{figure}
\includegraphics{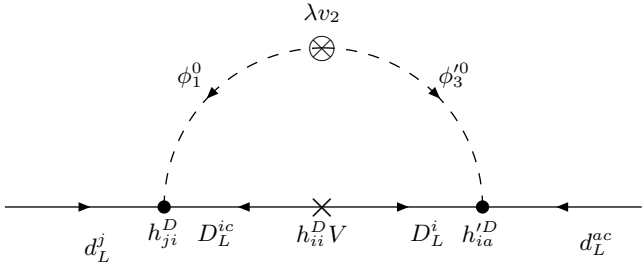}
\caption{\label{fig2}One loop diagram contributing to the radiative 
 generation of the down quark mass.}
\end{figure}

The ingredient that generates radiative masses for the two tree-level 
massless states is just the scalar field $\phi_3$ used in the up quark 
sector, via a Yukawa term of the form 
$Q^i_L\phi_3C\sum_ah^{\prime d}_{ia}d_L^{ac}$. The diagram in Fig.~\ref{fig2} shows how to generate these masses at the one loop 
level, one of them enhanced by sum of Yukawa couplings and the other one depleted by difference of Yukawa couplings.

Notice that the mass eigenstates are related to the weak eigenstates (up
to small mixing of the order of $v_1$ with the two exotic down quarks) in the following way: for the bottom quark $b\approx (d^3+d^2)/\sqrt{2}$, for the strange quark $s\approx (d^3-d^2)/\sqrt{2}$, and for the down quark
$d\approx d^1$ plus mixing via radiative loops with the other four
down type quarks. 

Our findings here are that the spectrum in the down quark sector is not as
neat as it is in the up quark sector, and that the up-down hierarchy is
unavoidable in the context of the analysis presented.

\subsection{\label{sec:sec33}The charged lepton sector}
The Yukawa term for the three charged leptons allowed by the discrete 
symmetry is
\begin{equation} 
{\cal L}_Y^l=\sum_{a,b=1}^3h_{ab}^eL_{aL}\phi_2^*Ce_{bL} + h.c., 
\end{equation}
which, in the basis $(e^1,e^2,e^3)$, produces a mass matrix of the form
\begin{equation}
M_e=v_2\left(\begin{array}{ccc}
h^e_{11} & h^e_{12} & h^e_{13} \\
h^e_{21} & h^e_{22} & h^e_{23} \\
h^e_{31} & h^e_{32}  & h^e_{33} \\
\end{array}\right).
\end{equation}
With all the Yukawa couplings equal, this is a democratic type mass matrix which, as we know, is a good starting point to generate consistent lepton mass matrices. The disadvantage here is again the strong hierarchy to be imposed between the Yukawa couplings in the up, down and leptonic sectors. 

We can avoid again the hierarchy by using a discrete symmetry in order
to eliminate the tree-level mass term, and introducing 
Higgses (leptoquarks) able to produce radiative masses for all the charged leptons.

\subsection{\label{sec:sec34}The neutral lepton sector} 
\begin{figure}
\includegraphics{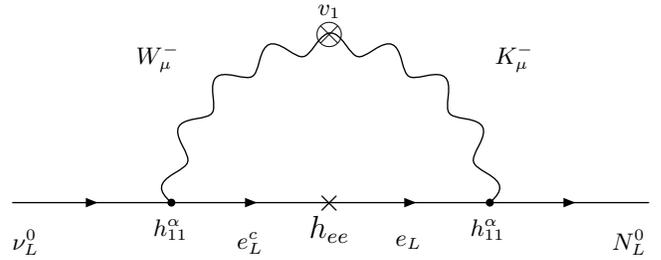}
\caption{\label{fig3}One loop diagram contributing to the radiative 
 generation of a Dirac mass for the neutrinos.}
\end{figure}

The discrete $Z_2$
symmetry avoids the only possible Yukawa term leading to tree-level neutrino masses \cite{kita1}, which is of the form ${\cal
L}_y^n=\lambda_{ab}L_L^aL_L^b\phi_2$. However, the neutrinos can get radiative Dirac type masses
using the mixing between the two charged gauge bosons as depicted in
Fig.~\ref{fig3}. The mass matrix generated in this way is of the democratic
type with large mixing among the several flavors. Notice also that the value of $v_1$ must be very small in order to have a small neutrino mass scale, in agreement with the analysis in Sec.~\ref{sec:sec2} (if $v_1=0$ the mechanism in Ref.~\cite{kita1} 
is still available).

Notice the particular way how the neutrinos mix and get masses in this
model. They acquire only Dirac type masses via a one loop radiative
correction, which is different to the well known ways of generating
neutrino masses, namely, the Zee mechanism \cite{zee} and the see-saw mechanism.

\section{\label{sec:sec4}Constraints on the gauge bosons mixing angles}
After the identification of the mass eigenstates, we can properly
bound $\sin\theta$ and $M_{Z_2}$ by using parameters measured at the $Z$
pole from CERN $e^+e^-$ collider (LEP), SLAC Linear Collider (SLC), and
atomic parity violation constraints which are given in Table~\ref{tab3}.

The expression for the partial decay width for $Z^{\mu}_1\rightarrow
f\bar{f}$ is
 
\begin{eqnarray}\nonumber
\Gamma(Z^{\mu}_1\rightarrow f\bar{f})&=&\frac{N_C G_F
M_{Z_1}^3}{6\pi\sqrt{2}}\rho \Big\{\frac{3\beta-\beta^3}{2}
[g(f)_{1V}]^2 \\ \label{ancho}
& + & \beta^3[g(f)_{1A}]^2 \Big\}(1+\delta_f)R_{EW}R_{QCD}, \quad
\end{eqnarray}
\noindent 
where $f$ is an ordinary SM fermion, $Z^\mu_1$ is the physical gauge boson
observed at LEP, $N_C=1$ for leptons while for quarks
$N_C=3(1+\alpha_s/\pi + 1.405\alpha_s^2/\pi^2 - 12.77\alpha_s^3/\pi^3)$,
where the 3 is due to color and the factor in parentheses represents the
universal part of the QCD corrections for massless quarks 
(for fermion mass effects and further QCD corrections which are 
different for vector and axial-vector partial widths, see 
Ref.~\cite{kuhn}); $R_{EW}$ are the electroweak corrections which include 
the leading order QED corrections given by $R_{QED}=1+3\alpha/(4\pi)$. 
$R_{QCD}$ are further QCD corrections (for a comprehensive review see 
Ref.~\cite{leike} and references therein), and $\beta=\sqrt{1-4 m_f^2/
M_{Z_1}^2}$ is a kinematic factor which can be taken equal to $1$ for all 
the SM fermions except for the bottom quark. 
The factor $\delta_f$ contains the one loop vertex
contribution which is negligible for all fermion fields except for the 
bottom quark, for which the contribution coming from the top quark, at the 
one loop vertex radiative correction, is parameterized as $\delta_b\approx 
10^{-2} [-m_t^2/(2 M_{Z_1}^2)+1/5]$~\cite{pich}. The $\rho$ parameter 
can be expanded as $\rho = 1+\delta\rho_0 + \delta\rho_V$ where the 
oblique correction $\delta\rho_0$ is given by
$\delta\rho_0\approx 3G_F m_t^2/(8\pi^2\sqrt{2})$, and $\delta\rho_V$ is 
the tree level contribution due to the $(Z_{\mu} - Z'_{\mu})$ mixing which 
can be parameterized as $\delta\rho_V\approx
(M_{Z_2}^2/M_{Z_1}^2-1)\sin^2\theta$. Finally, $g(f)_{1V}$ and $g(f)_{1A}$
are the coupling constants of the physical $Z_1^\mu$ field with ordinary
fermions which, for this model, are listed in Table~\ref{tab1}.

In what follows we are going to use the experimental values~\cite{pdb}:
$M_{Z_1}=91.188$ GeV, $m_t=174.3$ GeV, $\alpha_s(m_Z)=0.1192$,
$\alpha(m_Z)^{-1}=127.938$, and $\sin\theta^2_W=0.2333$. These
values are introduced using the definitions $R_\eta\equiv
\Gamma_Z(\eta\eta)/\Gamma_Z(hadrons)$ for $\eta=e,\mu,\tau,b,c,s,u,d$.

As a first result, notice from Table~\ref{tab1} that this model predicts 
$R_e=R_\mu=R_\tau$, in agreement with the experimental results in 
Table~\ref{tab3}, independent of any flavor mixing at the tree-level.

The effective weak charge in atomic parity violation, $Q_W$, can be 
expressed as a function of the number of protons $(Z)$ and the number of 
neutrons $(N)$ in the atomic nucleus in the form 

\begin{equation}
Q_W=-2\left[(2Z+N)c_{1u}+(Z+2N)c_{1d}\right], 
\end{equation}
\noindent
where $c_{1q}=2g(e)_{1A}g(q)_{1V}$. The theoretical value for $Q_W$ for 
the cesium atom is given by~\cite{ginges} $Q_W(^{133}_{55}Cs)=-73.19\pm 0.13 + \Delta Q_W$, where the contribution of new physics is included in $\Delta Q_W$ which can be written as~\cite{durkin}

\begin{equation}\label{DQ} 
\Delta 
Q_W=\left[\left(1+4\frac{S^4_W}{1-2S^2_W}\right)Z-N\right]\delta\rho_V
+\Delta Q^\prime_W.
\end{equation}

The term $\Delta Q^\prime_W$ is model dependent and it can be obtained for
our model by using $g(e)_{iA}$ and $g(q)_{iV}$, $i=1,2$, from Tables~\ref{tab1} and \ref{tab2}. The value we obtain is

\begin{equation}
\Delta Q_W^\prime=(3.75 Z + 2.56 N) \sin\theta + (1.22 Z + 0.41 N)
\frac{M^2_{Z_1}}{M^2_{Z_2}}\; .
\end{equation}

The discrepancy between the SM and the experimental data for $\Delta Q_W$ 
is given by~\cite{ginges}

\begin{equation}
\Delta Q_W=Q^{exp}_W-Q^{SM}_W=0.45\pm 0.48,
\end{equation}
which is $1.1\; \sigma$ away from the SM predictions.

\begin{table}
\caption{\label{tab3}Experimental data and SM values for some parameters 
related with neutral currents.}
\begin{ruledtabular}
\begin{tabular}{lcl}
& Experimental results & SM \\ \hline
$\Gamma_Z$(GeV)  & $2.4952 \pm 0.0023$  &  $2.4966 \pm 0.0016$  \\   
$\Gamma(had)$ (GeV)  & $1.7444 \pm 0.0020$ & $1.7429 \pm 0.0015$ \\ 
$\Gamma(l^+l^-)$ (MeV) & $83.984\pm 0.086$ & $84.019 \pm 0.027$ \\
$R_e$ & $20.804\pm 0.050$ & $20.744\pm 0.018$ \\ 
$R_\mu$ & $20.785\pm 0.033$ & $20.744\pm 0.018$ \\ 
$R_\tau$ & $20.764\pm 0.045$ & $20.790\pm 0.018$ \\ 
$R_b$ & $0.21664\pm 0.00068$ & $0.21569\pm 0.00016$ \\ 
$R_c$ & $0.1729\pm 0.0032$ & $0.17230\pm 0.00007$ \\ 
$Q_W^{Cs}$ & $-72.74\pm 0.29\pm 0.36$ & $-73.19\pm 0.13$  \\
$M_{Z_{1}}$(GeV) & $ 91.1872 \pm 0.0021 $ & $ 91.1870 \pm 0.0021 $ \\ 
\end{tabular}
\end{ruledtabular}
\end{table}

\begin{figure*}
\includegraphics{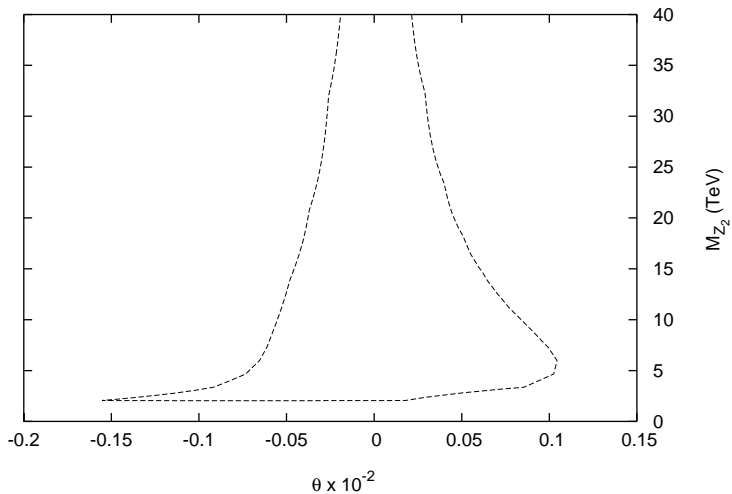}
\caption{\label{fig4}Contour plot displaying the allowed region for 
 $\theta$ vs $M_{Z_2}$ at 95\% C.L..}
\end{figure*}

Introducing the expressions for $Z$ pole observable in Eq.~(\ref{ancho}),
with $\Delta Q_W$ in terms of new physics in Eq.~(\ref{DQ}) and using
experimental data from LEP, SLC and atomic parity violation (see 
Table~\ref{tab3}), we do a $\chi^2$ fit and we find the best allowed region in the $(\theta-M_{Z_2})$ plane at $95\%$ confidence level (C.L.). 
In Fig.~\ref{fig4} we display this region which gives us the constraints
\begin{equation}\label{graph}
-0.00156\leq\theta\leq 0.00105, \quad \qquad 2.1\; {\mbox TeV} \leq
M_{Z_2}. 
\end{equation}
As we can see, the mass of the new neutral gauge boson is compatible with
the bound obtained in $p\bar{p}$ collisions at the Fermilab Tevatron
~\cite{abe}. From our analysis we can also see that for $\vert \theta \vert
\rightarrow 0$, $M_{Z_2}$ peaks at a finite value larger than $100$~TeV
which still copes with the experimental constraints on the $\rho$
parameter.

\section{\label{sec:sec5}Conclusions}
During the last decade several 3-3-1 models for one and three families
have been analyzed in the literature, the most popular one being the
Pleitez-Frampton model \cite{pf} which is certainly not the simplest
construction. Other two different three-family models, more appealing but not so popular
in the literature, were presented in Refs.~\cite{vl} and \cite{ozer}. The first one, studied in this paper, contains right-handed neutrinos, and
the second one without right-handed neutrinos but with an extra exotic
electron per family. The systematic analysis presented in Refs.~\cite{pfs} and \cite{pgs} shows that there are an infinite number of models based on the 3-3-1
gauge structure, most of them including particles with exotic electric
charges; but the number of models with particles without exotic electric
charges are just a few. Other two 3-3-1 models for one family and only with
ordinary electric charges are analyzed for example in Refs.~\cite{spm}.

For the model presented in this paper we have studied the mixing of the
two charged gauge bosons and its implications. Also we analyzed the mixing 
of the three electrically neutral gauge bosons and set limits for the 
mixing angles using precision measurements of the electroweak sector.

There are in this model three mass scales: the 3-3-1 scale $V\sim 1$ TeV,
the electroweak scale $v_2\approx 175$ GeV, and a new very small scale $v_1$ proportional to the neutrino mass scale

By the use of a discrete $Z_2$ symmetry we have constructed an appealing
mass spectrum for the fermions in this model; in particular we have
carried a program with a minimum set of Higgs scalars (three) and VEV in
which: the exotic fermions get heavy masses at the TeV scale; the quarks
$t,c$ and $d$ get tree-level masses, with the ordinary up quark sector getting
masses at the $v_2$ scale and with a hierarchy between the up and down sectors; the mass for the $c$ quark is suppressed by differences of Yukawa
couplings, and the other ordinary quarks, $b,s$ and $u$, get radiative masses, with the $Z_2$ symmetry responsible of the weak isospin breaking.

The quark mass spectrum constructed in this way is just one example of how 
to proceed. Different discrete symmetries and extra scalar Higgs fields with and without VEV 
may be used in order to avoid hierarchies and produce a more realistic mass 
spectrum \cite{fgp}.

The neutrino masses and oscillations have been analyzed for
this 3-3-1 model in Ref.~\cite{kita1} using a different approach from the one sketched here. In particular these authors use four $SU(3)_L$ scalar triplets instead of three, one of them with zero VEV as in this model, plus one charged scalar, three more $SU(3)_L$ singlets of neutral spin 1/2 particles (one per family), and the extra assumption of a diagonal mass matrix for the charged leptons. In the main text we have proposed an alternative way of generating tiny neutrino masses with large mixing, but a detailed analysis of our proposal is beyond the aims of this paper.

In summary, in this paper we have presented original results completing
previous analysis of the 3-3-1 model with right-handed neutrinos. First, the bounds on the mixing
parameters of the neutral currents have been updated. But most important,
our Higgs sector and VEV are different from the ones introduced in previous
papers. They imply quite different mass matrices for gauge bosons and
fermion fields, and a quite different phenomenology. The most
important fact about our Higgs sector is that it allows for an acceptable
fermion mass spectrum. Then we identify, for the first time for this model, the quark mass eigenstates; this allows us to do a
consistent phenomenological analysis and to set reliable bounds on new
physics coming from heavy neutral currents. It is worth noticing that the same scalar sector and VEV structure we have considered, have been used in Ref.~\cite{pr} in order to show that, by allowing for a nonzero and small VEV $v_1$, the spontaneous breaking of the lepton number can be implemented in the model. A new results to remark is also the embeding of this 3-3-1
structure into $SU(6)$.

The maybe unpleasant hierarchies present in the model can be avoided by
introducing more Higgs scalar fields and new VEV, but again, the analysis
is beyond the reach of this paper \cite{fgp}.

\section*{ACKNOWLEDGMENTS}
We acknowledge partial financial support from DIME at Universidad Nacional de Colombia-Sede Medell\'\i n, and from CODI at Universidad de Antioquia. W.A. Ponce thanks the INFN {\it laboratori di Frascati} in Italy for a pleasant
hospitality during the final stages of this work.  We thank E. Nardi 
and C. Garc\'\i a Canal for critical readings of the original manuscript and V. Pleitez for a written communication.

\end{document}